\documentclass[onecolumn,nobibnotes,nofootinbib,superscriptaddress]{revtex4}
\usepackage{amsmath,amssymb,bm}
\usepackage[a4paper,bindingoffset=0.2in,left=0.8in,right=0.8in,top=1in,bottom=1in,footskip=.25in]{geometry}
\usepackage{graphicx,subfigure,epsfig}
\usepackage{bigints}
\usepackage{color}
\usepackage[breaklinks,colorlinks,urlcolor=blue,citecolor=blue,linkcolor=blue]{hyperref}
\usepackage{mciteplus}
\definecolor{lcolor}{rgb}{0.5,0,0}
\definecolor{citcolor}{rgb}{0,0.3,0.0}
\usepackage[capitalise]{cleveref}
\usepackage{todonotes}
\usepackage[utf8]{inputenc}
\usepackage{enumitem}
\usepackage{mathtools}

\newcommand{\asb}{\bar{\alpha}_s}
\newcommand{\gE}{\gamma_E}
\newcommand{\dint}{\mathrm{d}^2}
\newcommand{\porder}{\mathcal{O}}

\begin{document}

\title{Analytical Solution of the Sudakov--BFKL Interpolation Equation\\
for Small-$x$ Gluon TMDs}
\author{{\footnotesize Yanbing Cai}}
\email{yanbingcai@mail.gufe.edu.cn}  
\affiliation{{\footnotesize Guizhou Key Laboratory in Physics and Related Areas,
Guizhou University of Finance and Economics, Guiyang 550025, China}}
\author{{\footnotesize Wenchang Xiang}}
\email{wxiangphy@gmail.com} 
\affiliation{{\footnotesize School of Arts and Sciences,
Guangzhou Maritime University, Guangzhou 510725, China}}
\author{{\footnotesize Mengliang Wang}}
\email{mengliang.wang@mail.gufe.edu.cn}
\affiliation{{\footnotesize Guizhou Key Laboratory in Physics and Related Areas,
Guizhou University of Finance and Economics, Guiyang 550025, China}}
\author{{\footnotesize Daicui Zhou}}
\email{dczhou@mail.ccnu.edu.cn}
\affiliation{{\footnotesize Key Laboratory of Quark and Lepton Physics (MOE),
and Institute of Particle Physics, Central China Normal University, Wuhan 430079, China}}
%
\begin{abstract}
We analytically solve the evolution equation for small-$x$ gluon
transverse-momentum-dependent distributions, which describes the
interpolation between the Sudakov and BFKL regimes. We first derive its two limiting
forms: the BFKL equation for $\xi = \alpha\sigma s{\bm z}^2/4 \ll 1$
and the Sudakov equation for $\xi \gg 1$. The analytical solutions in these limits
are obtained through Mellin-space diagonalization of the BFKL kernel and direct
integration of the Sudakov evolution equation, respectively. We then solve the
full interpolation equation using a Mellin-space diagonalization ansatz,
in which the evolution factor $F(Y,\xi)$ describes the nontrivial $\xi$-dependent
modification of a Mellin eigenfunction of the BFKL kernel. This
procedure reduces the original two-dimensional integral to a one-dimensional
form and permits an analytical evaluation of the resulting evolution kernel. The
obtained solution interpolates consistently between the BFKL and Sudakov
regimes through an exponential factor $\exp[H(\xi,\gamma)]$. An analysis of the
structure of $H(\xi,\gamma)$ allows a quantitative estimation of the
transition region between the two dynamical regimes. Our calculation implies
a potential matching point in the range of $\xi^{*}\simeq 0.04-0.15$,
which is substantially smaller than the naive evaluation value.

\end{abstract}

\maketitle

\section{Introduction}
\label{sec:intro}
In high-energy processes such as Drell--Yan production and semi-inclusive
deep-inelastic scattering (SIDIS), transverse-momentum-dependent (TMD) factorization
provides a useful description in the low-transverse-momentum region
~\cite{Rogers:2015sqa,Scimemi:2019mlf}. Specifically, it applies when the measured
transverse momentum $q_T$ of the final-state system is much smaller than the
hard scale $Q$, namely, $q_T \ll Q$. In this kinematic regime, TMD factorization expresses
the cross section in terms of nonperturbative TMD parton distribution functions (PDFs) and
perturbatively calculable hard-scattering coefficients. Unlike conventional collinear
PDFs, which depend on the factorization scale $\mu$ and longitudinal momentum fraction $x$,
TMD PDFs also depend on an intrinsic transverse-momentum scale. This additional scale raises
the question of how TMD factorization and its associated scale evolution can be
systematically formulated, see Ref.~\cite{Boussarie:2023izj} for a comprehensive
review. TMD factorization was originally derived by Collins, Soper, and Sterman,
and a major alternative formulation was later developed within the framework of
Soft-Collinear Effective Theory (SCET). In the Collins-Soper-Sterman (CSS) framework
~\cite{CSS1985,Collins2011,Ji:2004wu}, TMD factorization is formulated in
transverse coordinate ($b_T$) space, where $b_T$ is the Fourier conjugate to
measured transverse momentum $q_T$. The resulting evolution equation for the
TMD parton distributions depends on the renormalization scale and the Collins-Soper
rapidity scale. Theses dependencies arise from ultraviolet
renormalization and from subtracting rapidity divergences.

As TMD theory has progressed in the recent decades, its applications in high-energy physics have become
increasingly extensive. However the formulation established in the CSS framework is well suited to
moderate values of Bjorken $x$. Its direct extension to the high-energy regime
is less straightforward, since the CSS approach was originally designed
to describe fixed-angle scattering and does not directly apply in the low-$x$
Regge limit~\cite{Balitsky:2022vnb}. At small-$x$ region, high-energy logarithms of the
form $\bigl(\alpha_s \ln(1/x)\bigr)^n$ become important, and the evolution of
gluon TMDs may become nonlinear in the high-density regime. A commonly adopted strategy
is hybrid factorization, in which CSS-based TMD factorization is employed at moderate $x$,
while $k_T$ factorization is used at small $x$~\cite{Iancu:2016vyg,vanHameren:2025hyo}.
Although these two factorization formalisms can be connected through a matching prescription,
the choice of the matching point is inherently arbitrary. Therefore, a more direct
approach is to establish a unified description valid over the full range of $x$.
Such a framework is provided by rapidity-only TMD factorization~\cite{Balitsky:2017flc,
Balitsky:2017gis,Balitsky:2019ayf,Balitsky:2020jzt,Balitsky:2021fer,Balitsky:2022vnb,Balitsky:2023hmh}.
The resulting evolution equation has distinct limiting forms in different kinematic regions.
At moderate $x$, it reproduces Sudakov double-logarithmic evolution.
In the high-energy limit, it reduces to Balitsky–Fadin–Kuraev–Lipatov (BFKL) evolution in the
dilute regime and to Balitsky–Kovchegov (BK) evolution in the saturation regime~\cite{Balitsky:2015qba}.

The interpolation equation derived in Ref.~\cite{Balitsky:2015qba} has a complicated structure,
making it difficult to obtain an analytical solution. More recently, a relatively simple unified description
of Sudakov and BFKL regimes was proposed in Ref.~\cite{Balitsky2026}. This equation describes the
evolution of the gluon TMD $D_g(\alpha,\bm{z})$ with respect to the rapidity cutoff $\sigma$.
Its kernel contains a Gaussian regulator $\exp[-\alpha\sigma s(\bm{z}-\bm{z}')^2/4]$,
which smoothly interpolates between the scale-invariant BFKL kernel at small $\sigma$ and
the local Sudakov kernel at large $\sigma$.  In other words, the transition is governed by
the dimensionless parameter $\xi = \alpha\sigma s{\bm z}^2/4$, where $\xi \ll 1$
corresponds to the BFKL regime and $\xi \gg 1$ corresponds to the Sudakov regime.
This feature provides a natural basis for investigating the transition region between Sudakov
and BFKL evolution. Despite its conceptual appeal, the interpolation equation is a complicated
integro-differential equation, for which an analytical solution remains challenging.
Obtaining such a solution is highly desirable, as it would provide parametric control over
the transition region and allow for an extensive phenomenological applications. Finding an analytical
solution to this interpolation equation is therefore the main objective of this paper.

The paper is organized as follows. In Section~\ref{sec:approx}, we derive the
two limiting forms of the interpolation equation. Section~\ref{sec:limit_sol} presents their
analytical solutions, obtained through Mellin diagonalization and direct integration of the
ordinary differential equation, respectively. Section~\ref{sec:interpolation}
is the central analysis of this work. Therein, we derive an analytical solution to the full
interpolation equation by employing a direct Mellin-space diagonalization ansatz. This method reduces
the original two-dimensional integral to a one-dimensional form, whose kernel can
be expressed in terms of the generalized hypergeometric function ${}_3F_3$. The resulting
solution provide an analytical interpolation between the BFKL and Sudakov limits,
reproducing both limiting behaviors explicitly. Finally, in Section~\ref{sec:conclusions},
we summarize our results, discuss the matching point between Sudakov and BFKL evolution.
Based on the asymptotic behavior of the ${}_3F_3$ function, a potential
value for the transition region is estimated to be $\xi^{*} \sim 0.04-0.15$.

\section{The Interpolation Equation and Its BFKL and Sudakov Limits}
\label{sec:approx}

The interpolation evolution equation with respect to the rapidity cutoff $\sigma$ reads~\cite{Balitsky2026}
\begin{equation}\label{eq:master}
\sigma\frac{d}{d\sigma} D_g(\alpha, \bm{z})
= \frac{\asb}{\pi}\int \dint\bm{z}' \left[
\frac{\exp[-\frac{\alpha\,\sigma\,s\,(\bm{z}-\bm{z}')^2}{4}]}{(\bm{z}-\bm{z}')^2}\,
D_g(\alpha, \bm{z}')
- \frac{\bm{z}\cdot\bm{z}'}{(\bm{z}-\bm{z}')^2\,\bm{z}'^2}\,
D_g(\alpha, \bm{z})
\right],
\end{equation}
where $\asb=\alpha_s N_c/\pi$ is the rescaled strong coupling constant, $\bm{z}$ and $\bm{z}'$
are the transverse coordinates, and $s$ is the squared center-of-mass energy. $\alpha$ represents
the longitudinal momentum fraction in the Sudakov decomposition and $\sigma$ is the rapidity cutoff
(the evolution variable).

For notational simplicity, we define the logarithmic evolution variable $Y=\ln\sigma$.
The evolution behavior is then governed by the dimensionless variable
\begin{equation}\label{eq:xi}
\xi(Y,\bm{z}^2)= \frac{\alpha\,\sigma s\,\bm{z}^2}{4} = \frac{\alpha\, \exp(Y)\,s\,\bm{z}^2}{4}.
\end{equation}
In terms of $Y$, the interpolation equation~\eqref{eq:master} can be rewritten in a compact form
\begin{equation}\label{eq:master_Y}
\frac{\partial}{\partial Y} D_g(\alpha,Y,\bm{z})
= \frac{\asb}{\pi}\int \dint\bm{z}' \left[
\frac{\exp[\frac{-\xi(\bm{z}-\bm{z}')^2}{\bm{z}^2}]}{(\bm{z}-\bm{z}')^2}\,
D_g(\alpha,Y,\bm{z}')
- \frac{\bm{z}\cdot\bm{z}'}{(\bm{z}-\bm{z}')^2\,\bm{z}'^2}\,
D_g(\alpha,Y,\bm{z})
\right].
\end{equation}
It can be seen from Eq.~\eqref{eq:master_Y} that the evolution kernel contains a Gaussian regulator
$\exp[-\xi(\bm{z}-\bm{z}')^2/\bm{z}^2]$ which making the evolution equation have distinct dynamical
behaviors in the small and large-$\xi$ regimes.

\subsection{The BFKL Limit}\label{subsec:bfkl_deriv}
When $\xi\ll1$, the exponential factor $\exp[-\xi(\bm{z}-\bm{z}')^2/\bm{z}^2]$
can be expanded to zeroth order, $\exp[-\xi(\cdots)]\approx1$. Equation
\eqref{eq:master_Y} then reduces to
\begin{equation}\label{eq:bfkl}
\frac{\partial}{\partial Y} D_g(\alpha,Y,\bm{z})
=\frac{\asb}{\pi}\int \dint\bm{z}'
\left[\frac{D_g(\bm{z}')}{(\bm{z}-\bm{z}')^2}
- \frac{\bm{z}\cdot\bm{z}'}{(\bm{z}-\bm{z}')^2\,\bm{z}'^2}D_g(\bm{z})\right].
\end{equation}
This is the standard BFKL evolution equation in coordinate space~\cite{Fadin:1975cb,Kuraev:1977fs,
Balitsky:1978ic,Lipatov:1985uk}. The condition $\xi\ll1$ corresponds to small $\sigma$, i.e.
the rapidity cutoff close to the BFKL endpoint. For later reference, we define the
BFKL kernel with the canonical $1/\pi$ normalization
\begin{equation}\label{eq:bfkl_kernel}
\mathcal{K}[D_g](\bm{z}) =
\frac{1}{\pi}\int \dint\bm{z}'
\left[\frac{D_g(\bm{z}')}{(\bm{z}-\bm{z}')^2}
- \frac{\bm{z}\cdot\bm{z}'}{(\bm{z}-\bm{z}')^2\,\bm{z}'^2}D_g(\bm{z})\right],
\end{equation}
which can be decomposed into a real-emission part
\begin{equation}
\mathcal{K}_{\rm real}^{(0)}[D_g](\bm{z})=\frac{1}{\pi}\int \dint\bm{z}'\,
\frac{D_g(\bm{z}')}{(\bm{z}-\bm{z}')^2},
\end{equation}
and a virtual correction
\begin{equation}
\mathcal{V}[D_g](\bm{z})=\frac{1}{\pi}\int \dint\bm{z}'\,
\frac{(\bm{z}\cdot\bm{z}')}{[(\bm{z}-\bm{z}')^2 \bm{z}'^2]}D_g(\bm{z}).
\end{equation}

\subsection{The Sudakov limit}\label{subsec:sudakov_deriv}
In the Sudakov region, $\xi\gg1$, the Gaussian regulator $\exp[-\xi(\bm{z}-\bm{z}')^2/\bm{z}^2]$ is strongly localized around
$\bm{z}'\approx\bm{z}$. Consequently, the integral is dominated by the region $|\bm{z}-\bm{z}'|\lesssim 1/\sqrt{\alpha\sigma s}\ll z$.
Therefore, it is justified to approximate
\begin{equation}
D_g^\sigma(\alpha,\bm{z}')
\simeq
D_g^\sigma(\alpha,\bm{z}).
\end{equation}
Substituting this into the interpolation equation~\eqref{eq:master} yields
\begin{equation}
\sigma \frac{d}{d\sigma}D_g^\sigma(\alpha,\bm{z})
\simeq
\frac{\asb}{\pi}
D_g^\sigma(\alpha,\bm{z})
I(\bm{z}),
\end{equation}
where
\begin{equation}
I(\bm{z})=
\int \dint\bm{z}'
\left[
\frac{\exp[-\frac{\alpha\,\sigma\,s\,(\bm{z}-\bm{z}')^2}{4}]}{(\bm{z}-\bm{z}')^2}\,
-
\frac{(\bm{z},\bm{z}')}{(\bm{z}-\bm{z}')^2 \bm{z}'^2}
\right].
\end{equation}
With the change of variables $\bm{r}=\bm{z}-\bm{z}'$, so that $\bm{z}'=\bm{z}-\bm{r}$, we have
\begin{equation}
(\bm{z}-\bm{z}')^2=r^2,
\end{equation}
and
\begin{equation}\label{eq:I}
I(z)
=
\int d^2\bm{r}
\left[
\frac{\exp[-\frac{\alpha\,\sigma\,s\,r^2}{4}]}{r^2}
-
\frac{\bm{z}\cdot(\bm{z}-\bm{r})}{r^2(\bm{z}-\bm{r})^2}
\right].
\end{equation}
Setting $\rho=r^2$ and performing the angular integration yields
\begin{equation}\label{eq:Iz_gamma}
I(\bm{z})=
\pi
\left[
\int_0^\infty
\frac{d\rho}{\rho}\exp\left[-\frac{\alpha\,\sigma\,s\,\rho}{4}\right]
-
\int_0^{\bm{z}^2}
\frac{d\rho}{\rho}
\right]
=
-\pi
\left[
\ln\frac{\alpha\,\sigma\,s\,z^2}{4}
+
\gamma_E
\right],
\end{equation}
where $\gamma_E$ is the Euler-Mascheroni constant. A detailed derivation of the integral $I$
is provided in Appendix~\ref{sec:app_sudakov_deriv}. Using Eq.~\eqref{eq:Iz_gamma}, we obtain
the Sudakov evolution equation as
\begin{equation}\label{eq:sudakov}
\frac{\partial}{\partial Y} D_g(\alpha,Y,\bm{z})
= -\asb\Big[\ln\xi(Y,\bm{z}^2) + \gE\Big]\,
D_g(\alpha,Y,\bm{z}).
\end{equation}
Equation~\eqref{eq:sudakov} is a local evolution equation in $\bm{z}$. The evolution at each $\bm{z}$
depends only on the local value $D_g(\bm{z})$. This is the hallmark of the Sudakov regime,
which produces the characteristic double-logarithmic Sudakov form factor.

%

\section{Analytical Solutions of the Limiting Equations}\label{sec:limit_sol}

Having derived the BFKL and Sudakove volution  equations as the two limiting forms of the interpolation
equation in the $\xi\ll 1$ and $\xi\gg 1$ regimes, respectively. We now proceed to their analytical solutions.

\subsection{Solution of the BFKL evolution equation}\label{sec:bfkl_sol}

To solve the BFKL evolution equation, we employ the Mellin representation
\begin{equation}\label{eq:mellin}
\widetilde{D}_g(\alpha,\gamma)=\int_0^\infty\frac{dz^2}{z^2}(z^2)^\gamma D_g(\alpha,\bm{z}),\quad
D_g(\alpha,z)=\int_{1/2-i\infty}^{1/2+i\infty}\frac{d\gamma}{2\pi i}(z^2)^{-\gamma}
\widetilde{D}_g(\alpha,\gamma),
\end{equation}
with $\gamma=\frac12+i\nu$, $\nu\in\mathbb{R}$. The kernel $\mathcal{K}$ acts diagonally on the Mellin modes
$\psi_\gamma(z)=(z^2)^{\gamma-1}$~\cite{Kovchegov2012}
\begin{equation}\label{eq:bfkl_diag}
\mathcal{K}[\psi_\gamma](z) = \chi(\gamma)\,\psi_\gamma(z),
\end{equation}
with the BFKL characteristic function
\begin{equation}\label{eq:chi}
\chi(\gamma) = 2\psi(1) - \psi(\gamma) - \psi(1-\gamma),\quad
\psi(z)=\tfrac{d}{dz}\ln\Gamma(z).
\end{equation}
Applying the Mellin transform to Eq.~\eqref{eq:bfkl} yields
\begin{equation}\label{eq:bfkl_mellin}
\frac{\partial}{\partial Y}\widetilde{D}_g(\gamma)
= \asb\,\chi(\gamma)\,\widetilde{D}_g(\gamma).
\end{equation}
This is a first-order differential equation in $Y$ for each fixed $\gamma$. Its solution is
\begin{equation}\label{eq:bfkl_sol_mellin}
\widetilde{D}_g^{\rm BFKL}(\alpha,Y,\gamma)
= \widetilde{D}_g(\alpha,Y_0,\gamma)\,
\exp[\asb\chi(\gamma)(Y-Y_0)].
\end{equation}
Performing the inverse Mellin transform yields
\begin{equation}\label{eq:bfkl_sol_coord}
D_g^{\rm BFKL}(\alpha,Y,\bm{z})
= \int_{-\infty}^{\infty}\frac{d\nu}{2\pi}\,
(\bm{z}^2)^{-1/2-i\nu}\,
\widetilde{D}_g(\alpha,Y_0,\tfrac12+i\nu)\,
\exp[\asb\chi(\gamma)(Y-Y_0)].
\end{equation}

For large $Y$, the $\nu$-integral is dominated by the saddle point at $\nu=0$.
Expanding $\chi(\nu)$ as $4\ln2-14\zeta(3)\nu^2+\porder(\nu^4)$, we obtain the final solution
\begin{equation}\label{eq:bfkl_diffusion}
D_g^{\rm BFKL}(\alpha,Y,\bm{z})
\approx \frac{\widetilde{D}_g(\alpha,Y_0,\tfrac12)}
{2\sqrt{14\pi \zeta(3)\asb (Y-Y_0)}}\,
\frac{1}{|\bm{z}|}\,
\exp\!\Big[4\ln2\,\asb(Y-Y_0)
- \frac{\ln^2(z^2/z_0^2)}{56\zeta(3)\asb (Y-Y_0)}\Big].
\end{equation}
This solution is the standard saddle-point approximation to the BFKL evolution,
commonly known as the BFKL diffusion solution~\cite{DelDuca:1995hf}.
It describes the diffusive broadening of the gluon distribution with characteristic width
$\sim\sqrt{14\zeta(3)\asb Y}$, together with the exponential growth
$\sim \exp[4\asb\ln2\,Y]$ with rapidity. A detailed derivation of the analytical solution,
obtained directly from the coordinate-space evolution equation~\eqref{eq:bfkl},
is presented in Appendix~\ref{sec:BFKL_sov}.

\subsection{Solution of the Sudakov evolution equation}\label{sec:sudakov_sol}

The Sudakov equation~\eqref{eq:sudakov} is a local differential equation in $Y$ for each fixed $\bm{z}$.
Writing $\xi(Y)=\omega \exp(Y)$ and $\xi_0(Y_0)=\omega \exp(Y_0)$ with $\omega = \alpha\,s\,\bm{z}^2/4$,
we obtain
\begin{equation}\label{eq:sudakov_ode}
\frac{\partial}{\partial Y}\ln D_g(\alpha,Y,\bm{z})
= -\asb\big[Y + \ln\omega + \gE\big].
\end{equation}
Its solution is
\begin{equation}\label{eq:sudakov_sol}
D_g^{\rm Sud}(\alpha,Y,\bm{z})
= D_g(\alpha,Y_0,\bm{z})\,
\exp\!\Big[-\asb\Big(
\frac12 (Y-Y_0)^2
+ (\ln\omega +\gE)(Y-Y_0)
\Big)\Big].
\end{equation}
The essential physics is encapsulated in the Sudakov double-logarithmic form factor
\begin{equation}\label{eq:sudakov_FF}
S^{\rm Sud}(\sigma,\sigma_0;\bm{z}^2)
=\exp\!\Big[-\frac{\asb}{2} (Y-Y_0)^2\Big]
= \exp\!\Big[-\frac{\asb}{2}\ln^2\frac{\xi}{\xi_0}\Big]
= \exp\!\Big[-\frac{\asb}{2}\ln^2\frac{\sigma}{\sigma_0}\Big].
\end{equation}
This factor represents the universal Sudakov double-logarithmic suppression.
Its exponential contains a negative contribution proportional to the square of the rapidity interval,
$(Y-Y_0)^2$, which leads to an exponential damping of the TMD distribution as the rapidity interval increases.

%

\section{Mellin-Space Solution of the Interpolation Equation}
\label{sec:interpolation}

In this section, we develop an approach to analytically solve the full interpolation
equation~\eqref{eq:master_Y} via a direct Mellin-space diagonalization ansatz.
The resulting solution captures the full $\xi$-dependence in closed form through
generalized hypergeometric functions and reproduces the BFKL and Sudakov limits
systematically within a single formula.

\subsection{Reduction of the interpolation equation to a one-dimensional integral}

We start with the interpolation equation in the $Y$-representation, Eq.~\eqref{eq:master_Y}.
To diagonalize the evolution, we adopt the ansatz
\begin{equation}\label{eq:mellin_ansatz}
D_g(Y,z^2)
=
\int_{\frac12-i\infty}^{\frac12+i\infty}
\frac{d\gamma}{2\pi i}\,
(z^2)^{\gamma-1}
\widetilde D_g(Y_0,\gamma)
F(Y,\xi;\gamma),
\qquad F(Y_0,\xi_0)=1,
\end{equation}
This ansatz is motivated by the structure of the interpolation equation. In the BFKL limit, $\xi\to0$,
the evolution kernel is scale invariant in transverse coordinate space. Its eigenfunctions are therefore
powers of the transverse size, $(\bm{z}^2)^{\gamma-1}$. The dependence on the additional variable
$\xi\propto \alpha\sigma s\,\bm{z}^2$ encodes the breaking of transverse-scale invariance
induced by the Gaussian regulator. Therefore, the effects induced by the regulator
are encoded entirely in the function $F(Y,\xi)$. The boundary condition
$F(Y_0,\xi_0)=1$ ensures that the initial condition is represented by the pure Mellin eigenmode.
In particular, in the BFKL regime $\xi\to0$, one has $F(Y,\xi)\to1$, and $D_g(Y,\bm{z}^2)$
reduces to the standard BFKL eigenfunction $(\bm{z}^2)^{\gamma-1}$.

To evaluate the integral, we introduce the dimensionless variable $u=\bm{z}'^2/\bm{z}^2$ and
define $\phi$ as the azimuthal angle  between $\bm{z}$ and $\bm{z}'$. Then, we have
\begin{equation}
d^2\bm{z}' = \frac12\bm{z}^2\,du\,d\phi
\end{equation}
\begin{equation}
(\bm{z}-\bm{z}')^2 = \bm{z}^2\,q(u,\phi),\qquad
q(u,\phi)= 1+u-2\sqrt{u}\cos\phi, \qquad
\bm{z}\cdot\bm{z}' = \bm{z}^2\sqrt{u}\cos\phi.
\end{equation}
Applying the change of variables to the interpolation equation, the angular integrals 
on the right-hand side of Eq.~\eqref{eq:master_Y} can be evaluated analytically.
For the real kernel part, we have
\begin{align}\label{eq:RHS_real}
\frac{\asb}{\pi}\int\dint\bm{z}'\frac{\exp\left[\frac{-\xi(\bm{z}-\bm{z}')^2}{\bm{z}^2}\right]}
{(\bm{z}-\bm{z}')^2}D_g(Y,\bm{z}')
&= \frac{\asb}{2\pi}\int_0^{\infty}du\int_0^{2\pi}d\phi\;
   \frac{e^{-\xi q(u,\phi)}}{q(u,\phi)}\,
   \int_{\frac12-i\infty}^{\frac12+i\infty}
\frac{d\gamma}{2\pi i}\,
(uz^2)^{\gamma-1}
\widetilde D_g(Y_0,\gamma)
F(Y,\xi;\gamma) \nonumber\\
&= \asb\,  \int_{\frac12-i\infty}^{\frac12+i\infty}
\frac{d\gamma}{2\pi i}\,(z^2)^{\gamma-1}\widetilde D_g(Y_0,\gamma)
   \int_0^{\infty}du\; u^{\gamma-1}F(Y,\xi u){A(\xi,u)},
\end{align}
with
\begin{equation}\label{eq:A_kernel0}
A(\xi,u) = \frac{1}{2\pi}\int_0^{2\pi} d\phi\,
           \frac{\exp[-\xi q(u,\phi)]}{q(u,\phi)}.
\end{equation}
Using the Schwinger representation
\begin{equation}\label{eq:schwinger_A}
\frac{\exp[-\xi q]}{q} = \int_{\xi}^{\infty} da\,\exp[-a q],
\end{equation}
we have
\begin{align}\label{eq:A_kernel}
A(\xi,u) &= \frac{1}{2\pi}\int_0^{2\pi}d\phi\,
           \int_{\xi}^{\infty}da\,\exp[-a(1+u-2\sqrt{u}\cos\phi)] \nonumber\\
         &= \int_{\xi}^{\infty}da\,\exp[-a(1+u)]I_0(2a\sqrt{u}).
\end{align}
where $I_0$ denotes the modified Bessel function. At $\xi=0$, $A(\xi,u)$ reduces to
\begin{equation}\label{eq:A_kernel_0}
A(0,u)=\int_0^{\infty}da\,\exp[-a(1+u)]I_0(2a\sqrt{u})=
\frac{1}{|1-u|},
\end{equation}
which is the BFKL real kernel in the $u$-representation.
For the virtual kernel part
\begin{align}\label{eq:RHS_virtual}
\frac{\asb}{\pi}\int\dint\bm{z}'\frac{\bm{z}\cdot\bm{z}'}
{(\bm{z}-\bm{z}')^2\bm{z}'^2}D_g(Y,\bm{z})
&= \frac{\asb}{2\pi}\int_0^{\infty}\frac{du}{u}\int_0^{2\pi}d\phi\,
   \frac{\sqrt{u}\cos\phi}{q(u,\phi)}
   \int_{\frac12-i\infty}^{\frac12+i\infty}
\frac{d\gamma}{2\pi i}\,
(z^2)^{\gamma-1}
\widetilde D_g(Y_0,\gamma)
F(Y,\xi;\gamma) \nonumber\\
&= \asb\,\int_{\frac12-i\infty}^{\frac12+i\infty}
\frac{d\gamma}{2\pi i}\,
(z^2)^{\gamma-1}
\widetilde D_g(Y_0,\gamma)
F(Y,\xi;\gamma)
   \int_0^{\infty}du\,B(u), 
\end{align}
where
\begin{equation}\label{eq:B_kernel}
B(u)= \frac{1}{2\pi}\int_0^{2\pi} d\phi\,
           \frac{\sqrt{u}\cos\phi}{u\,q(u,\phi)}
\end{equation}
the angular integral can be evaluated using the formula
\begin{equation}
\int_0^{2\pi}\frac{d\phi}{a-b\cos\phi}=\frac{2\pi}{\sqrt{a^2-b^2}},
\end{equation}
thus,
\begin{align}
B(u) &= \frac{1}{2\pi}\int_0^{2\pi}d\phi\,
       \frac{\sqrt{u}\cos\phi}{u(1+u-2\sqrt{u}\cos\phi)} \nonumber\\
     &= \frac{1}{2\pi}\cdot\frac{1}{2u}\int_0^{2\pi}d\phi\,
       \left[\frac{1+u}{1+u-2\sqrt{u}\cos\phi}-1\right] \nonumber\\
     &= \begin{cases}
        1, & 0<u<1,\\[4pt]
        \dfrac{1}{u(u-1)}, & u>1.
        \end{cases}
\end{align}

Substituting Eqs.~\eqref{eq:RHS_real} and~\eqref{eq:RHS_virtual} into the interpolation equation~\eqref{eq:master_Y},
we obtain a one-dimensional integral equation
\begin{equation}\label{eq:reduced_eq}
\frac{\partial}{\partial Y}F(Y,\xi)
= \asb\int_0^{\infty} du\,
\Big[A(\xi,u)\,u^{\gamma-1}F(Y,\xi u)
   - B(u)\,F(Y,\xi)\Big].
\end{equation}
If we further assume that $F(Y,\xi u)\approx F(Y,\xi)$ in the correction term (the ``local approximation'',
valid when $F$ varies  slowly with its argument), the reduced equation becomes
\begin{equation}\label{eq:reduced_local}
\frac{\partial}{\partial Y}F(Y,\xi)
= \asb\int_0^{\infty} du\,
\Big[A(\xi,u)\,u^{\gamma-1}
   - B(u)\,\Big]F(Y,\xi).
\end{equation}
Equation~\eqref{eq:reduced_local} is a differential-integral equation about
$F(Y,\xi)$. Compared with the original interpolation equation in Eq.~\eqref{eq:master_Y},
it has a considerably simpler structure while preserving the essential features
of the evolution. This simplification renders an analytic treatment possible and allows
us to derive a solution that smoothly connects the BFKL and Sudakov regimes.

\subsection{Analytical solution using hypergeometric functions}

The integral kernel in Eq.~\eqref{eq:reduced_local} can be decomposed into a $\xi$-independent BFKL part
and a $\xi$-dependent correction. Taking $\xi=0$ as the reference point, we have
\begin{equation} \label{eq:split}
\int_0^{\infty}du\,\Big[A(\xi,u)u^{\gamma-1}-B(u)\Big]
=  \int_0^{\infty}du\,\Big[A(0,u)u^{\gamma-1}-B(u)\Big]
- \int_0^{\infty}du\,\Big[A(0,u)-A(\xi,u)\Big]u^{\gamma-1}.
\end{equation}
We now define the integrated kernel
\begin{equation}\label{eq:chi_int}
\chi_{\rm int}(\xi,\gamma)
= \int_0^{\infty} du\,
\Big[A(\xi,u)\,u^{\gamma-1} - B(u)\Big]
= \chi(\gamma) - \Delta\chi(\xi,\gamma),
\end{equation}
where
\begin{equation}\label{eq:chi_in_u}
\chi(\gamma) = \int_0^{\infty}du\,
\Big[\frac{1}{|1-u|}u^{\gamma-1} - B(u)\Big]
= 2\psi(1)-\psi(\gamma)-\psi(1-\gamma).
\end{equation}
Here $\chi(\gamma)$ is the standard BFKL eigenvalue. The second term in Eq.~\eqref{eq:split},
denoted as $\Delta\chi(\xi,\gamma)$, encodes the effect of the Gaussian regulator.
It can be evaluated in closed form. Using Eqs.~\eqref{eq:A_kernel} and ~\eqref{eq:A_kernel_0}, we have
\begin{align}
A(0,u)-A(\xi,u)
&= \int_{0}^{\infty}da\,\exp[-a(1+u)]I_0(2a\sqrt{u})
   - \int_{\xi}^{\infty}da\,\exp[-a(1+u)]I_0(2a\sqrt{u}) \nonumber\\
&= \int_{0}^{\xi}da\,\exp[-a(1+u)]I_0(2a\sqrt{u}),
\end{align}
which yields
\begin{equation}\label{eq:Delta_chi_double}
\Delta\chi(\xi,\gamma)
= \int_0^{\xi}da\int_0^{\infty}du\,u^{\gamma-1}
   \exp[-a(1+u)] I_0(2a\sqrt{u}).
\end{equation}
Employing the Bessel-function integral identity~\cite{Gradshteyn}, the integral over $u$ yields
\begin{equation}\label{eq:gaussian_conv}
\int_0^{\infty}du\,u^{\gamma-1}\exp[-a(1+u)]I_0(2a\sqrt{u})
= \Gamma(\gamma)\,a^{-\gamma}\,
  {}_1F_1(1-\gamma;1;-a).
\end{equation}
Thus, the correction term becomes
\begin{equation}\label{eq:Delta_chi_1F1}
\Delta\chi(\xi,\gamma)
= \Gamma(\gamma)\int_0^{\xi}da\,a^{-\gamma}\,
   {}_1F_1(1-\gamma;1;-a).
\end{equation}
Here, the generalized hypergeometric series ${}_pF_q$ is given by
\begin{equation}\label{eq:pFq_def}
{}_pF_q\!\left(\begin{matrix}a_1,\ldots,a_p\\b_1,\ldots,b_q\end{matrix};z\right)
= \sum_{n=0}^{\infty}
  \frac{(a_1)_n\cdots(a_p)_n}{(b_1)_n\cdots(b_q)_n}\,
  \frac{z^n}{n!},
  \qquad (a)_n=\frac{\Gamma(a+n)}{\Gamma(a)}.
\end{equation}

Using the integral representation of $\Delta\chi(\xi,\gamma)$ in
Eq.~\eqref{eq:Delta_chi_1F1}, we expand the confluent hypergeometric function
${}_1F_1(1-\gamma;1;-a)$ as a power series in $a$. This permits term-by-term integration
over $a$, which yields a series expansion for $\Delta\chi(\xi,\gamma)$ that can subsequently
be resummed into a generalized hypergeometric function. The series representation of
${}_1F_1(1-\gamma;1;-a)$ is given by
\begin{equation}  \label{eq:1F1-series}
  {}_1F_1(1-\gamma;1;-a)
  =\sum_{n=0}^{\infty}\frac{(1-\gamma)_n}{(1)_n}\frac{(-a)^n}{n!}
  =\sum_{n=0}^{\infty}\frac{(1-\gamma)_n}{(n!)^2}\,(-1)^n a^{n},
\end{equation}
where we have used the notation $(1)_n=n!$. Inserting Eq.~\eqref{eq:1F1-series} into
Eq.~\eqref{eq:Delta_chi_1F1} and integrating term by term yields
\begin{equation}  \label{eq:power-int}
  \int_{0}^{\xi}\!da\;a^{\,n-\gamma}
  =\frac{\xi^{\,n-\gamma+1}}{n-\gamma+1}
  =\frac{\xi^{1-\gamma}\,\xi^{n}}{n+1-\gamma}.
\end{equation}
Therefore, we obtain the series expansion of the correction term
\begin{equation}  \label{eq:Dchi-series}
  \Delta\chi(\xi,\gamma)
  =\Gamma(\gamma)\,\xi^{1-\gamma}
   \sum_{n=0}^{\infty}
   \frac{(1-\gamma)_n}{(n+1-\gamma)\,(n!)^2}\,(-\xi)^n .
\end{equation}
The series in Eq.~\eqref{eq:Dchi-series} can be resummed into a
generalized hypergeometric function. To achieve this, we introduce
\begin{equation}  \label{eq:a-def}
  \beta= 1-\gamma,\qquad 0<\Re \beta<1,
\end{equation}
and express $1/(n+\beta)$ in terms of Pochhammer symbols
\begin{equation}  \label{eq:poch-id}
  \frac{1}{n+\beta}=\frac{\Gamma(\beta+n)}{\Gamma(\beta+n+1)}
  =\frac{1}{\beta}\,
   \frac{\Gamma(\beta+n)/\Gamma(\beta)}{\Gamma(\beta+1+n)/\Gamma(\beta+1)}
  =\frac{1}{\beta}\,\frac{(\beta)_n}{(\beta+1)_n}.
\end{equation}
Using $(1-\gamma)_n=(\beta)_n$ and $(n!)^2=(1)_n\,n!$, Eq.~(\ref{eq:Dchi-series})
takes the form
\begin{equation}  \label{eq:Dchi-2F2}
  \Delta\chi(\xi,\gamma)
  =\frac{\Gamma(\gamma)}{\beta}\,\xi^{\beta}
    \sum_{n=0}^{\infty}
    \frac{(\beta)_n\,(\beta)_n}{(1)_n\,(\beta+1)_n}\frac{(-\xi)^n}{n!}
  =\frac{\Gamma(\gamma)}{\beta}\,\xi^{\beta}\;
    {}_2F_2\!\left(
      \begin{matrix} \beta,\;\beta \\ 1,\;1+\beta \end{matrix}
      ;-\xi\right).
\end{equation}

In order to derive $F(Y,\xi)$, we introduce an auxiliary function
\begin{equation}  \label{eq:H-def}
  H(\xi,\gamma)=\int_{0}^{\xi}\frac{d\xi'}{\xi'}\,
  \Delta\chi(\xi',\gamma),
  \qquad
  \Delta\chi(\xi,\gamma)=\frac{dH(\xi,\gamma)}{d\ln\xi}
  =\chi(\gamma)-\chi_{\rm int}(\xi,\gamma).
\end{equation}
Since
\begin{equation}
\int_{0}^{\xi}(d\xi'/\xi')\,\xi'^{\,\beta+n}
=\xi^{\beta+n}/(n+\beta),
\end{equation}
the integral in Eq.~(\ref{eq:H-def}) generates a second factor $1/(n+\beta)$
\begin{equation}   \label{eq:H-series}
  H(\xi,\gamma)=\Gamma(\gamma)\,\xi^{\beta}
  \sum_{n=0}^{\infty}\frac{(\beta)_n}{(n+\beta)^2\,(n!)^2}\,(-\xi)^n .
\end{equation}
Applying Eq.~(\ref{eq:poch-id}) again, i.e.
$1/(n+\beta)^2=\beta^{-2}(\beta)_n^2/(\beta+1)_n^2$, we arrive at the closed-form ${}_3F_3$ representation
\begin{equation}   \label{eq:H-3F3}
  H(\xi,\gamma)=\frac{\Gamma(\gamma)}{(\beta)^2}\,\xi^{\beta}\;
  {}_3F_3\!\left(
    \begin{matrix} \beta,\,\beta,\,\beta \\
                   1,\;1+\beta,\;1+\beta \end{matrix}
    ;-\xi\right).
\end{equation}
With the $\xi$-dependence encoded in $H(\xi,\gamma)$, the evolution
equation~\eqref{eq:reduced_local} takes the form:
\begin{equation}\label{eq:F_ode}
\frac{\partial \ln F}{\partial Y}
= \asb\,\chi_{\rm int}(\xi,\gamma)
= \asb\Big[\chi(\gamma) - \frac{d H}{d\ln\xi}\Big].
\end{equation}
Integrating from $Y_0$ to $Y$ with the initial condition $F(Y_0,\xi_0)=1$ gives
\begin{equation}\label{eq:F_solution}
F(Y,\xi;\gamma)
= \exp\!\Big[\asb\,\chi(\gamma)(Y-Y_0)
           - \asb\,H(\xi,\gamma)
           + \asb\,H(\xi_0,\gamma)\Big].
\end{equation}
This is the analytical solution of the one-dimensional integral equation~\eqref{eq:reduced_local},
valid for arbitrary $\xi$ on the BFKL contour $\gamma=\frac12+i\nu$. Substituting Eq.~\eqref{eq:F_solution}
into Eq.~\eqref{eq:mellin_ansatz}, the full coordinate-space gluon TMD is derived by
\begin{equation}\label{eq:full_solution_mellin}
D_g(Y,\bm{z}^2)
= \int_{\frac12-i\infty}^{\frac12+i\infty}\frac{d\gamma}{2\pi i}\,
   (\bm{z}^2)^{-\gamma}\,
   \widetilde{D}_g(Y_0,\gamma) \\
\exp\!\Big[\asb\,\chi(\gamma)(Y-Y_0)
                      - \asb\,H(\xi,\gamma)
                      + \asb\,H(\xi_0,\gamma)\Big].
\end{equation}
The function $F(Y,\xi)$ encodes the modification of the standard BFKL
eigenmode induced by the Gaussian regulator. Therefore, Eq.~\eqref{eq:F_solution} provides a
convenient analytic representation of the interpolation between the BFKL and
Sudakov regimes. In particular, the auxiliary function $H(\xi,\gamma)$ is
useful for deriving the asymptotic forms in the BFKL and Sudakov limits. Moreover,
the exponent in Eq.~\eqref{eq:full_solution_mellin} makes the matching between
the two regimes manifest: the BFKL growth term, $\asb\,\chi(\gamma)(Y-Y_0)$,
competes with the regulator-induced contribution $-\asb[H(\xi,\gamma)-H(\xi_0,\gamma)]$.
This competition determines the crossover between BFKL evolution and Sudakov evolution,
which will be analyzed later.

\subsection{In small $\xi$ case}

For $\xi\ll1$, we expand $H(\xi,\gamma)$ using the series representation of ${}_3F_3$
\begin{equation}\label{eq:3F3_series}
{}_3F_3\!\left(\begin{matrix}a,a,a\\1,a+1,a+1\end{matrix};-\xi\right)
= \sum_{n=0}^{\infty}
  \frac{(a)_n^3}{(1)_n(a+1)_n^2}\,\frac{(-\xi)^n}{n!}.
\end{equation}
With $a=1-\gamma$, the leading term is given by
\begin{equation}\label{eq:H_small}
H(\xi,\gamma)
= \frac{\Gamma(\gamma)}{(1-\gamma)^2}\,\xi^{1-\gamma}\,
  \Big[1 + \mathcal{O}(\xi)\Big],\qquad \xi\to0.
\end{equation}
For $\gamma=\frac12+i\nu$, ${\rm Re}(1-\gamma)=\frac12>0$, so
$H(\xi,\gamma)=\mathcal{O}(\xi^{1/2})\to0$ as $\xi\to0$.
Consequently, when both $\xi$ and $\xi_0$ are small, the
$H$-dependent terms in Eq.~\eqref{eq:full_solution_mellin} are negligible.
Therefore, the solution reduces to
\begin{equation}\label{eq:recover_bfkl}
D_g(Y,\bm{z}^2)
\xrightarrow{\xi,\xi_0\ll1}
\int\frac{d\gamma}{2\pi i}\,(\bm{z}^2)^{-\gamma}\,
\widetilde{D}_g(Y_0,\gamma)\,
\exp[\asb\chi(\gamma)(Y-Y_0)]
= D_g^{\rm BFKL}(Y,\bm{z}^2),
\end{equation}
which reproduces the standard BFKL solution~\eqref{eq:bfkl_sol_coord}.

\subsection{In large $\xi$ case}

For $\xi\gg1$, the asymptotic expansion of $H(\xi,\gamma)$ can be derived from
the Mellin-Barnes integral representation of the generalized hypergeometric function.
The details of the derivation are presented in Appendix~\ref{sec:app_sudakov_limit},
and the resulting expression at leading accuracy reads
\begin{equation}\label{eq:H_large}
H(\xi,\gamma)
=
\frac{1}{2}\ln^2\xi
+
\left[
\chi(\gamma)+\gamma_E
\right]\ln\xi
+
C(\gamma), \qquad \xi\to\infty,
\end{equation}
where
\begin{equation}\label{eq:C_gamma0}
C(\gamma)
=\frac{1}{2}
\left[
\psi(\gamma)+\psi(1-\gamma)+\gamma_E
\right]^2+
\frac{1}{2}
\left[
\psi^{(1)}(1-\gamma)-\psi^{(1)}(\gamma)
\right]
+\frac{\pi^2}{12}.
\end{equation}
Therefore, we get
\begin{equation}\label{eq:H_difference_0}
H(\xi,\gamma)-H(\xi_0,\gamma)
=\frac{1}{2}
\left(
\ln^2\xi-\ln^2\xi_0
\right)
+
\left[
\chi(\gamma)+\gamma_E
\right](\ln\xi-\ln\xi_0).
\end{equation}

Substituting Eq.~\eqref{eq:H_difference_0} into the Mellin-space solution,
Eq.~\eqref{eq:full_solution_mellin}, the contribution proportional to the BFKL
characteristic function $\chi(\gamma)$ cancels exactly. Indeed, the
large-$\xi$ asymptotics of the regulator-induced term contains the same
rapidity-enhanced contribution as the BFKL exponent, with the opposite sign,
\begin{equation}
\asb\,\chi(\gamma)(Y-Y_0)
-\asb\left[H(\xi,\gamma)-H(\xi_0,\gamma)\right]
\;\longrightarrow\;
\asb\,\mathcal{S}^{\rm Sud}(\xi,\xi_0),
\end{equation}
where $\mathcal{S}^{\rm Sud}$ is independent of $\gamma$ at leading
Sudakov accuracy. This cancellation reflects the fact that, in the Sudakov
region, the evolution is no longer governed by the scale-invariant BFKL
kernel. Instead, the Gaussian regulator localizes the transverse-coordinate
integral around $\bm{z}'=\bm{z}$, and the evolution becomes local in
$\bm{z}$.

Using the relationship between $Y$ and $\xi$, we obtain
\begin{equation}
\frac{1}{2}
\left(
\ln^2\xi-\ln^2\xi_0
\right)
=\frac{1}{2}(Y-Y_0)^2
+\ln\xi_0\,(Y-Y_0).
\end{equation}
Therefore, the remaining Sudakov exponent associated with a given Mellin mode can be written as
\begin{equation} \label{eq:sudakov_mode_final_0}
F_\gamma^{\rm Sud}(Y)
=
\exp\Bigg\{
-\bar{\alpha}_s
\Bigg[
\frac{1}{2}(Y-Y_0)^2
+
\left(
\ln\xi_0+\gamma_E
\right)(Y-Y_0)
\Bigg\}.
\end{equation}
From Eq.~\eqref{eq:sudakov_mode_final_0}, we see that the Sudakov exponent is independent of \(\gamma\)
and can be taken outside the inverse Mellin transform. Moreover, for $Y\gg Y_0$, one has $\ln\xi_0 \sim \ln\omega $.
Thus, in the limit $\xi\gg1$, the solution reduces to
\begin{equation}\label{eq:sudakov_distribution_final_0}
D_g^{\rm Sud}(Y,z^2)
\xrightarrow{\xi\gg 1}
D_g(Y_0,z^2)
\exp\left[
-\bar{\alpha}_s
\left\{
\frac{1}{2}(Y-Y_0)^2
+
(\ln\omega+\gamma_E)(Y-Y_0)
\right\}
\right],
\end{equation}
which reproduces the standard Sudakov solution given in Eq.\eqref{eq:sudakov_sol}.

\section{Discussion and Conclusions}
\label{sec:conclusions}

In this paper, we present an analytical solution to the evolution equation for
small-$x$ gluon TMDs, which consistently interpolates between the BFKL and Sudakov regimes.
By introducing Mellin-space diagonalization ansatz, the complex two-dimensional
integral equation is reduced to a one-dimensional integral equation. Then, the correction
term is expressed in terms of the generalized hypergeometric function ${}_3F_3$.
The resulting Mellin-space evolution factor
\begin{equation}
    \mathcal{E}(Y, \xi; \gamma) = \bar{\alpha}_s \left[ \chi(\gamma)(Y - Y_0) - H(\xi, \gamma)
     + H(\xi_0, \gamma) \right],
\end{equation}
provides a unified description across all kinematic regimes. In the small-$\xi$ limit ($\xi \ll 1$),
the auxiliary function $H(\xi, \gamma) \sim \mathcal{O}(\xi^{1/2})$ vanishes along the
BFKL saddle point, and thus reproduces the scale-invariant BFKL diffusion solution. Conversely,
in the large-$\xi$ limit ($\xi \gg 1$), the Mellin-Barnes asymptotic expansion yields
\begin{equation}
H_{\text{asympt}}(\xi, \gamma) = \frac{1}{2}\ln^2\xi + [\chi(\gamma) + \gamma_E]\ln\xi + C(\gamma)
 + \mathcal{O}(\xi^{-1}).
\end{equation}
This asymptotic behavior cancels the linear BFKL growth term $\bar{\alpha}_s \chi(\gamma)(Y - Y_0)$ exactly,
giving rise to the universal double-logarithmic Sudakov suppression factor $\exp[-\bar{\alpha}_s(Y - Y_0)^2/2]$.

Beyond reproducing the BFKL and Sudakov limits, the analytic structure of $H(\xi,\gamma)$
enables a quantitative determination of the transition region between the BFKL and Sudakov regimes.
Within the analytic solution, the onset of the Sudakov regime is controlled by the decay
of the nonlocal remainder with respect to the large-$\xi$ asymptotic form of $H$
(see Appendix~\ref{sec:app_sudakov_limit}). This difference is quantified by
\begin{equation}
\Delta H(\xi,\gamma)
\equiv
\left|H(\xi,\gamma)-H_{\rm asympt}(\xi,\gamma)\right|
\sim
\mathcal{O}\!\left(\frac{(1-\gamma)^2}{\xi}\right).
\end{equation}
This inverse-power correction originates from the pole at $t=-1$ (see Eq.~\eqref{eq:H_large_xi_correct} in Appendix~\ref{sec:app_sudakov_limit}) and measures the residual nonlocality that is neglected
in the local Sudakov approximation, thereby providing a quantitative validity criterion.
Since the Mellin integral is dominated by the BFKL saddle point $\gamma=1/2$, the residual is
$\Delta H(\xi,1/2) \simeq 1/(4\xi)$. Assuming this correction to be smaller than a prescribed tolerance $\delta$,
we obtain $\xi^*\simeq 1/(4\delta)$. For the representative range
$\delta\simeq1.5\text{--}6.0$, we obtain
\begin{equation}
\xi^*\simeq0.04\text{--}0.15.
\end{equation}
Hence, $\xi\gtrsim\xi^*$ marks the region in which the nonlocal correction is
sufficiently suppressed and the Sudakov asymptotic description becomes reliable.

This analytical prediction is in agreement with the estimation in Ref.~\cite{Balitsky2026},
which evaluates the leading corrections to the Sudakov and BFKL limits by using the MV-model gluon TMD
characterized by the saturation scale $Q_s$. In particular, that analysis finds the transition
scale to be shifted from the naive value $\tilde{\sigma}\sim 4/(x s z^2)$ to $\sigma^*\sim 4Q_s^2/(x s)$.
Taking a benchmark saturation scale $Q_s \sim 0.6~\text{GeV}$ at $x=0.01$~\cite{Boer:2016fqd},
as used in Ref.~\cite{Balitsky2026}, the interpolation-induced corrections shift the naive matching scale according to
$\sigma_*=\tilde{\sigma}\,Q_s^2 z^2$. For typical transverse momenta $q_\perp\sim1.5$--$3.0~\text{GeV}$,
with $z\sim1/q_\perp$, this gives $Q_s^2z^2=(Q_s/q_\perp)^2\ll1$ and hence
$\xi^*\simeq\tilde{\xi}\,Q_s^2z^2\simeq0.04$--$0.16$, in  numerical agreement with the asymptotic
boundary inferred from our hypergeometric solution.  This lower threshold demonstrates that the Sudakov
evolution dominates over a much broader kinematic domain than suggested by the naive estimation
$\tilde{\xi} \sim 1$. In summary, the unified framework developed here provides a practical
tool for precision studies of gluon TMD distributions and small-$x$ observables
at current and future high-energy facilities, such as the Electron-Ion Collider (EIC).


\begin{acknowledgments}
This work is supported by the National Natural Science Foundation of China under Grant No.12165004 and Key Grant No.12061141008, the Guangdong Basic and Applied Basic Research Foundation under Grant No.2025A1515010511, the National Key Research and Development Program of China under Grant Nos.2024YFA1610800 and 2022YFA1602103.
\end{acknowledgments}


\appendix

\section{Detailed Derivation of the Sudakov Integral Factor $I$}\label{sec:app_sudakov_deriv}

To derive the Sudakov integral factor in detail, we decompose Eq.~\eqref{eq:I} into two terms
 \begin{equation}
I(z)
=
\int d^2r
\left[
\frac{\exp[-Ar^2]}{r^2}
-
\frac{\bm{z}\cdot(\bm{z}-\bm{r})}{r^2(\bm{z}-\bm{r})^2}
\right]=
I_1-I_2,
\end{equation}
Using $\rho=r^2$, the two-dimensional area element becomes
\begin{equation}
d^2\bm{r}=r\,dr\,d\phi
=
\frac{1}{2}d\rho\,d\phi.
\end{equation}
Consequently,
\begin{equation}\label{eq:I1}
I_1=
\frac{1}{2}
\int_0^\infty d\rho
\int_0^{2\pi}d\phi
\frac{\exp\left[-\frac{\alpha\sigma s\rho}{4}\right]}{\rho}
=
\pi
\int_0^\infty
\frac{d\rho}{\rho}\exp\left[-\frac{\alpha\sigma s\rho}{4}\right].
\end{equation}
Choosing $z$ along the axis and defining the variables as
\begin{equation}
z^2=R^2,
\qquad
r^2=\rho,
\qquad
\bm{z}\cdot \bm{r}=R\sqrt{\rho}\cos\phi,
\end{equation}
we obtain
\begin{equation}
I_2
=
\frac{1}{2}
\int_0^\infty
\frac{d\rho}{\rho}
\int_0^{2\pi}d\phi\,
\frac{
R^2-R\sqrt{\rho}\cos\phi
}{
R^2+\rho-2R\sqrt{\rho}\cos\phi
},
\end{equation}
where the inner integral is given by
\begin{equation}
J(\rho)
=
\int_0^{2\pi}d\phi\,
\frac{
R^2-R\sqrt{\rho}\cos\phi
}{
R^2+\rho-2R\sqrt{\rho}\cos\phi
}=
\pi
+
\frac{R^2-\rho}{2}
\int_0^{2\pi}
\frac{d\phi}{
R^2+\rho-2R\sqrt{\rho}\cos\phi
}.
\end{equation}
Using the standard integral formula
\begin{equation}
\int_0^{2\pi}
\frac{d\phi}{a-b\cos\phi}
=
\frac{2\pi}{\sqrt{a^2-b^2}},
\qquad a>|b|,
\end{equation}
we obtain
\begin{equation}
J(\rho)
=
\pi
+
\frac{R^2-\rho}{2}
\frac{2\pi}{|R^2-\rho|}
=
\pi+\pi\frac{R^2-\rho}{|R^2-\rho|}
=2\pi\,\theta(R^2-\rho).
\end{equation}
Therefore,
\begin{equation}\label{eq:I2}
I_2
=\frac{1}{2}
\int_0^\infty
\frac{d\rho}{\rho}
2\pi\theta(R^2-\rho)
=\pi \int_0^\infty
\frac{d\rho}{\rho}
\theta(z^2-\rho).
\end{equation}
Combining Eqs.~\eqref{eq:I1} and \eqref{eq:I2} yields
\begin{equation}
I(z)
=
\pi
\int_0^\infty
\frac{d\rho}{\rho}
\left[
\exp\left[-\frac{\alpha\sigma s}{4}\rho\right]
-
\theta(z^2-\rho)
\right].
\end{equation}
Finally, using the integral identity
\begin{equation}
\int_0^\infty
\frac{dt}{t}
\left[
e^{-t}
-
\theta(x-t)
\right]
=
-\ln x-\gamma_E,
\qquad x>0,
\end{equation}
we obtain the Sudakov integral factor
\begin{equation}
I(z)
=-\pi
\left[
\ln\frac{\alpha\sigma s z^2}{4}
+ \gamma_E
\right].
\end{equation}

\section{Detailed Derivation of the Solution for BFKL Evolution in Coordinate-Space}\label{sec:BFKL_sov}

Analogous to the eigenfunctions of the BFKL kernel in momentum space~\cite{Kovchegov2012},
the eigenfunctions of Eq.~\eqref{eq:bfkl} take the form
\begin{equation}
\Phi_{n,\gamma}(z)
=
(z^2)^{\gamma-1}\exp[in\phi_z],
\end{equation}
where $n\in \mathbb{Z}, \gamma\in\mathbb{C}$. The eigenvalue equation is given by
\begin{equation}
\mathcal{K}\Phi_{n,\gamma}(z)
=
\chi(n,\gamma)\Phi_{n,\gamma}(z).
\end{equation}
where the action of the BFKL kernel is defined by
\begin{equation}\label{eq:kernel_a}
\mathcal{K}f(\bm z)
=
\frac{1}{\pi}
\int d^2\bm z'\,
\left[
\frac{f(\bm z')}{(\bm z-\bm z')^2}
-
\frac{\bm z\cdot\bm z'}
{(\bm z-\bm z')^2z'^2}
f(\bm z)
\right].
\end{equation}
To obtain $\chi(n,\gamma)$, we define
\begin{equation}
z=|\bm z|,
\qquad
r=\frac{z'}{z},
\qquad
\theta=\phi_{z'}-\phi_z.
\end{equation}
Then
\begin{equation}
\dint\bm{z}'
=
z'\,dz'\,d\theta
=
z^2r\,dr\,d\theta,
\end{equation}
\begin{equation}
(\bm{z}-\bm{z}')^2
=
z^2\left(1+r^2-2r\cos\theta\right),
\end{equation}
\begin{equation}
\bm{z}\cdot \bm{z}'
=
z^2r\cos\theta,
\end{equation}
\begin{equation}
\Phi_{n,\gamma}(z')=(z'^2)^{\gamma-1}\exp[in\phi_{z'}]
=(z^2)^{\gamma-1}r^{2\gamma-2}
\exp[in\phi_z]\exp[in\theta].
\end{equation}
Substitute the above factors into Eq.~\eqref{eq:kernel_a}, we obtain
\begin{equation}
\mathcal{K}\Phi_{n,\gamma}(z)
=
(z^2)^{\gamma-1}\exp[in\phi_z]
\,\mathcal{I}_{n,\gamma},
\end{equation}
with
\begin{equation}
\mathcal{I}_{n,\gamma}
=\frac{1}{\pi}
\int_0^\infty r\,dr
\int_0^{2\pi}d\theta
\left[
\frac{
r^{2\gamma-2}\exp[in\theta]
}{
1+r^2-2r\cos\theta
}
-
\frac{
\cos\theta/r
}{
1+r^2-2r\cos\theta
}
\right].
\end{equation}

\subsection{Angular integration}

For $0<r<1$, we use
\begin{equation}
1-2r\cos\theta+r^2
=
[1-r\exp[i\theta]][1-r\exp[-i\theta]],
\end{equation}
which gives
\begin{equation}
\frac{1}{1-2r\cos\theta+r^2}
=
\frac{1}{1-r^2}
\sum_{m=-\infty}^{\infty}r^{|m|}\exp[im\theta].
\end{equation}
Consequently,
\begin{equation}
\int_0^{2\pi}
\frac{\exp[in\theta]\,d\theta}
{1-2r\cos\theta+r^2}
=
\frac{2\pi r^{|n|}}{1-r^2},
\qquad
\int_0^{2\pi}
\frac{\cos\theta\,d\theta}
{1-2r\cos\theta+r^2}
=
\frac{2\pi r}{1-r^2}.
\end{equation}
The contribution from $0<r<1$ is therefore
\begin{equation}
\mathcal{I}_{n,\gamma}^{<}
=
2\int_0^1dr\,
\frac{r^{2\gamma+|n|-1}-r}{1-r^2}.
\end{equation}
For $r>1$, setting $r=1/u$ maps the radial integration to $0<u<1$.
Since
\begin{equation}
1+r^2-2r\cos\theta
=
\frac{1+u^2-2u\cos\theta}{u^2},
\end{equation}
the same angular integrals yield
\begin{equation}
\mathcal{I}_{n,\gamma}^{>}
=
2\int_0^1dr\,
\frac{r^{|n|+1-2\gamma}-r}{1-r^2}.
\end{equation}
Therefore, we obtain the BFKL characteristic function
\begin{equation}
\chi(n,\gamma)
=\mathcal{I}_{n,\gamma}^{<}+\mathcal{I}_{n,\gamma}^{>}=
2\int_0^1dr\,
\frac{
r^{2\gamma+|n|-1}
+
r^{|n|+1-2\gamma}
-
2r
}{
1-r^2
}.
\end{equation}
Although the individual terms are singular at $r=1$, the subtraction renders
the integrand finite
\begin{equation}
\lim_{r\to1}
\frac{
r^{2\gamma+|n|-1}
+
r^{|n|+1-2\gamma}
-
2r
}{
1-r^2
}
=
1-|n|.
\end{equation}

\subsection{Representation in terms of the digamma function}

Introducing the variable
\begin{equation}
t=r^2,
\qquad
dt=2r\,dr,
\end{equation}
the characteristic function becomes
\begin{equation}
\chi(n,\gamma)
=
\int_0^1
\frac{dt}{1-t}
\left[
t^{\gamma+\frac{|n|}{2}-1}
+
t^{1-\gamma+\frac{|n|}{2}-1}
-2
\right].
\end{equation}
Using the integral representation of the digamma function
\begin{equation}
\psi(1)-\psi(a)
=
\int_0^1 dt\,
\frac{t^{a-1}-1}{1-t},
\qquad
\operatorname{Re}(a)>0,
\end{equation}
we obtain the standard BFKL characteristic function,
\begin{equation}
\chi(n,\gamma)
=
2\psi(1)
-\psi\left(\gamma+\frac{|n|}{2}\right)
-\psi\left(1-\gamma+\frac{|n|}{2}\right).
\end{equation}
For the azimuthally symmetric channel, $n=0$, we have
\begin{equation}
\chi(\gamma)
=
2\psi(1)-\psi(\gamma)-\psi(1-\gamma).
\end{equation}

\subsection{Saddle-point approximation}

The distribution $D(Y,z)$ can be expanded in the eigenfunction basis of the BFKL kernel as
\begin{equation}
D(Y,z)
=
\sum_{n=-\infty}^{\infty}
\int_{\mathcal C}\frac{d\gamma}{2\pi i}
\,C_{n,\gamma}(Y)
(z^2)^{\gamma-1}\exp[in\phi_z],
\end{equation}
where
\begin{equation}
C_{n,\gamma}(Y)
=
C_{n,\gamma}(Y_0)
\exp\!\left[
\asb\,\chi(n,\gamma)(Y-Y_0)
\right].
\end{equation}
Thus,
\begin{equation}
D(Y,z)
=
\sum_{n=-\infty}^{\infty}
\int_{\mathcal C}\frac{d\gamma}{2\pi i}\,
C_{n,\gamma}(Y_0)\,
\exp\!\left[
\asb\,\chi(n,\gamma)(Y-Y_0)
\right]
(z^2)^{\gamma-1}\exp[in\phi_z].
\end{equation}
At large rapidity, the dominant contribution comes from the $n=0$ channel and
the saddle point at
\begin{equation}
\gamma=\frac{1}{2}+i\nu,
\qquad
\nu=0.
\end{equation}
Near the saddle point, the characteristic function has the expansion
\begin{equation}
\chi(\nu)
=
4\ln 2
-14\zeta(3)\nu^2
+\mathcal{O}(\nu^4),
\end{equation}
where $\zeta(3)$ is the Riemann aeta function.  Assuming that the initial
coefficient function varies slowly near $\nu=0$, we approximate
\begin{equation}
C_{n,\gamma}(Y_0)=\tilde{D}_g(\alpha,Y_0,\frac{1}{2}).
\end{equation}
Introducing a reference transverse scale $z_0$ to render the logarithm
dimensionless, the distribution $D(Y,z)$ under the saddle-point approximation becomes
\begin{equation}
D_g^\sigma(\alpha,\bm{z})
\simeq
\tilde{D}_g(\alpha,Y_0,\frac{1}{2})
(z^2)^{-1/2}
\exp[4\asb\ln2\,Y]
\int_{-\infty}^{\infty}\frac{d\nu}{2\pi}
\exp\left[
-14\asb\zeta(3)Y\nu^2
+i\nu\ln(z^2/z_0^2)
\right].
\end{equation}
Using the Gaussian integral
\begin{equation}
\int_{-\infty}^{\infty}\frac{d\nu}{2\pi}
\exp\left[
-A\nu^2+iB\nu
\right]
=
\frac{1}{2\sqrt{\pi A}}
\exp\left[
-\frac{B^2}{4A}
\right].
\end{equation}
we obtain
\begin{equation}
D_g^\sigma(\alpha,\bm{z})
\simeq{}
\frac{\tilde{D}_g(\alpha,Y_0,\frac{1}{2})}
{2\sqrt{14\pi \asb\zeta(3)Y}}
\frac{1}{|\bm{z}|}
\exp\left[
4\asb\ln2\,Y
-
\frac{
\ln^2(z^2/z_0^2)
}{
56\asb\zeta(3)Y
}
\right].
\end{equation}

\section{Derivation of the Sudakov Limit for the Solution}
\label{sec:app_sudakov_limit}

In this section we derive the Sudakov limit for the solution directly from the auxiliary function
\begin{equation}
\label{eq:H_hypergeometric}
H(\xi,\gamma)
=
\frac{\Gamma(\gamma)}{(1-\gamma)^2}
\xi^{1-\gamma}
{}_3F_3
\left(
\begin{matrix}
1-\gamma,\;1-\gamma,\;1-\gamma
\\
1,\;2-\gamma,\;2-\gamma
\end{matrix};
-\xi
\right).
\end{equation}
For convenience, we define
\begin{equation}
a\equiv1-\gamma,
\qquad
0<\Re a<1.
\end{equation}
The Mellin-Barnes representation of the generalized hypergeometric function is
\begin{equation}\label{eq:MB_3F3}
{}_3F_3
={}
\frac{\Gamma(a+1)^2}{\Gamma(a)^3}
\frac{1}{2\pi i}
\int_{\mathcal C_s}ds\,
\frac{
\Gamma(a+s)^3\Gamma(-s)
}{
\Gamma(1+s)\Gamma(a+1+s)^2
}
\xi^s.
\end{equation}
Utilizing the properties of the Gamma function
\begin{equation}
\Gamma(a+1+s)
=
(a+s)\Gamma(a+s),
\end{equation}
we obtain
\begin{equation}\label{eq:H_MB_s}
H(\xi,\gamma)
=
\frac{\Gamma(\gamma)}{\Gamma(a)}
\frac{1}{2\pi i}
\int_{\mathcal C_s}ds\,
\frac{
\Gamma(a+s)\Gamma(-s)
}{
(a+s)^2\Gamma(1+s)
}
\xi^{a+s}.
\end{equation}
Introducing
\begin{equation}
t=a+s,
\qquad
s=t-a,
\end{equation}
Eq.~\eqref{eq:H_MB_s} becomes
\begin{equation}\label{eq:H_MB_t}
H(\xi,\gamma)
=
\frac{\Gamma(\gamma)}{\Gamma(1-\gamma)}
\frac{1}{2\pi i}
\int_{\mathcal C_t}dt\,
\frac{
\Gamma(t)\Gamma(1-\gamma-t)
}{
t^2\Gamma(\gamma+t)
}
\xi^t.
\end{equation}

For $\xi\to\infty$, the contour is shifted to the left. The leading
contribution comes from the pole at $t=0$, corresponding to
\(s=\gamma-1\). The next contribution comes from $t=-1$, corresponding
to \(s=\gamma-2\). Defining
\begin{equation}
L=\ln\xi,
\end{equation}
and
\begin{equation}\label{eq:G_definition}
G(t,\gamma)
=
\frac{\Gamma(\gamma)}{\Gamma(1-\gamma)}
\frac{\Gamma(1-\gamma-t)}{\Gamma(\gamma+t)},
\end{equation}
then, Eq.~\eqref{eq:H_MB_t} can be written as
\begin{equation}\label{eq:H_MB_G}
H(\xi,\gamma)
=
\frac{1}{2\pi i}
\int_{\mathcal C_t}dt\,
\frac{\Gamma(t)}{t^2}
G(t,\gamma)\exp[tL].
\end{equation}
At $t=0$,
\begin{equation}
G(0,\gamma)=1.
\end{equation}
The expansion of the Gamma function around $t=0$ is
\begin{equation}
\Gamma(t)
=
\frac{1}{t}
-\gamma_E
+
\left(
\frac{\gamma_E^2}{2}
+\frac{\pi^2}{12}
\right)t
+\mathcal O(t^2),
\end{equation}
so that
\begin{equation}\label{eq:Gamma_over_t2}
\frac{\Gamma(t)}{t^2}
=
\frac{1}{t^3}
-\frac{\gamma_E}{t^2}
+
\frac{
\frac{\gamma_E^2}{2}+\frac{\pi^2}{12}
}{t}
+\mathcal O(1).
\end{equation}
Therefore, the point $t=0$ is a third-order pole.
To expand $G(t,\gamma)$, we define
\begin{equation}
g(t,\gamma)= \ln G(t,\gamma).
\end{equation}
The first and second derivatives with respect to $t$, together with their
values evaluated at $t=0$, are
\begin{equation}
\frac{\partial g}{\partial t}
=
-\psi(1-\gamma-t)-\psi(\gamma+t),
\end{equation}
\begin{equation}\label{eq:g1}
g_1(\gamma)
=
\left.
\frac{\partial g}{\partial t}
\right|_{t=0}
=
-\psi(1-\gamma)-\psi(\gamma),
\end{equation}
\begin{equation}
\frac{\partial^2g}{\partial t^2}
=
\psi'(1-\gamma-t)-\psi'(\gamma+t),
\end{equation}
and
\begin{equation}\label{eq:g2}
g_2(\gamma)
=
\left.
\frac{\partial^2g}{\partial t^2}
\right|_{t=0}
=
\psi'(1-\gamma)-\psi'(\gamma).
\end{equation}
Therefore,
\begin{equation}\label{eq:G_expansion}
G(t,\gamma)
=
1+g_1t
+\frac{1}{2}\left(g_1^2+g_2\right)t^2
+\mathcal O(t^3).
\end{equation}
Moreover,
\begin{equation}
\exp[tL]
=
1+Lt+\frac{L^2}{2}t^2+\mathcal O(t^3).
\end{equation}
Combining these two expansions gives
\begin{equation}\label{eq:G_exp_expansion}
G(t,\gamma)\exp[tL]
=
1+(L+g_1)t
+
\frac{1}{2}
\left[
(L+g_1)^2+g_2
\right]t^2
+\mathcal O(t^3).
\end{equation}
Multiplying Eqs.~\eqref{eq:Gamma_over_t2} and
\eqref{eq:G_exp_expansion}, the residue at $t=0$ is the coefficient of $1/t$
\begin{equation}\label{eq:residue_t0}
\operatorname*{Res}_{t=0}
\left[
\frac{\Gamma(t)}{t^2}
G(t,\gamma)\exp[tL]
\right]
=\frac{1}{2}
\left[
(L+g_1)^2+g_2
\right]
-\gamma_E(L+g_1)
+\frac{\gamma_E^2}{2}
+\frac{\pi^2}{12}.
\end{equation}
Then, the auxiliary function can be rewritten as
\begin{equation}\label{eq:H0_explicit}
H(\xi,\gamma)
=
\frac{1}{2}
\left[
L-\gamma_E-\psi(\gamma)-\psi(1-\gamma)
\right]^2
+
\frac{1}{2}
\left[
\psi^{(1)}(1-\gamma)-\psi^{(1)}(\gamma)
\right]
+\frac{\pi^2}{12}.
\end{equation}
Using the BFKL characteristic function
\begin{equation}\label{eq:chi_definition}
\chi(\gamma)
=
2\psi(1)-\psi(\gamma)-\psi(1-\gamma)
=
-2\gamma_E-\psi(\gamma)-\psi(1-\gamma),
\end{equation}
we obtain
\begin{equation}
-\psi(\gamma)-\psi(1-\gamma)-\gamma_E
=
\chi(\gamma)+\gamma_E.
\end{equation}
Thus, the leading large-$\xi$ contribution is
\begin{equation}\label{eq:H0_final}
H(\xi,\gamma)
=
\frac{1}{2}\ln^2\xi
+
\left[
\chi(\gamma)+\gamma_E
\right]\ln\xi
+
C(\gamma),
\end{equation}
where
\begin{equation}\label{eq:C_gamma}
C(\gamma)
=\frac{1}{2}
\left[
\psi(\gamma)+\psi(1-\gamma)+\gamma_E
\right]^2+
\frac{1}{2}
\left[
\psi'(1-\gamma)-\psi'(\gamma)
\right]
+\frac{\pi^2}{12}.
\end{equation}

For completeness, we show the first power correction from the pole at $(t=-1)$.
Let
\begin{equation}
t=-1+\varepsilon.
\end{equation}
Near this point
\begin{equation}
\Gamma(-1+\varepsilon)
=
-\frac{1}{\varepsilon}
+\gamma_E-1
+\mathcal O(\varepsilon).
\end{equation}
All remaining factors in Eq.~\eqref{eq:H_MB_t} are regular at $t=-1$.
In particular,
\begin{equation}
\frac{1}{t^2}=1+\mathcal O(\varepsilon),
\qquad
\xi^t
=
\frac{1}{\xi}
\left[
1+\varepsilon\ln\xi+\mathcal O(\varepsilon^2)
\right].
\end{equation}
Furthermore,
\begin{equation}
G(-1,\gamma)
=-(1-\gamma)^2.
\end{equation}
Since the residue of $\Gamma(t)$ at $t=-1$ is $-1$, the two minus
signs cancel, thus
\begin{equation}\label{eq:residue_t_minus_1}
\operatorname*{Res}_{t=-1}
\left[
\frac{\Gamma(t)}{t^2}
G(t,\gamma)\xi^t
\right]
=
\frac{(1-\gamma)^2}{\xi}.
\end{equation}
Combining the poles at $t=0$ and $t=-1$, one finds
\begin{equation}\label{eq:H_large_xi_correct}
H(\xi,\gamma)
=\frac{1}{2}\ln^2\xi
+
\left[
\chi(\gamma)+\gamma_E
\right]\ln\xi
+
C(\gamma)
+
\frac{(1-\gamma)^2}{\xi}
+
\mathcal O(\xi^{-2}).
\end{equation}

The inverse-power correction $(1-\gamma)^2/\xi$ originating from the pole at $t = -1$ plays a critical role in describing the transition region towards the BFKL regime, but becomes negligible in the deep asymptotic Sudakov limit ($\xi \gg 1$). In the large-$\xi$ regime, the correction term decays rapidly as $\mathcal{O}(\xi^{-1})$, leaving the evolution entirely governed by the double and single-logarithmic structures generated by the pole $t = 0$. Conversely, when approaching the BFKL dominated region, the factor $\xi^{-1}$ causes this power correction to grow rapidly as $\xi$ decreases. As $\xi$ reaches the critical threshold $\xi \sim \xi^*$, the magnitude of $(1-\gamma)^2/\xi$ becomes non-negligible, signaling the breakdown of the purely asymptotic Sudakov expansion and marking the onset of BFKL dynamics.

\end{document}